\documentclass[12pt,preprint]{aastex}

\shorttitle{CO Observations of LCBGs}
\shortauthors{Garland et al.}

\begin{document}

\title{The Nature of Nearby Counterparts to Intermediate Redshift Luminous Compact
Blue Galaxies \\ II.  CO Observations}

\author{C. A. Garland\footnotemark[1]~\footnotemark[2]~,
J. P. Williams\footnotemark[1]~,
D. J. Pisano\footnotemark[3]~\footnotemark[4]~\footnotemark[5],
R. Guzm\'an\footnotemark[6]~,
F. J. Castander\footnotemark[7]~,
and J. Brinkmann\footnotemark[8]}

\footnotetext[1]{Institute for Astronomy, University of Hawai`i, 2680
Woodlawn Drive, Honolulu, HI 96822;
catherine.garland@castleton.edu, jpw@ifa.hawaii.edu}
\footnotetext[2]{Present address: Natural Sciences Department, 
Castleton State College, Castleton, VT 05735}
\footnotetext[3]{CSIRO Australia Telescope National Facility, P.O. Box
76, Epping, NSW 1710, Australia; DJ.Pisano@csiro.au}
\footnotetext[4]{Bolton Fellow \& NSF MPS Distinguished International 
Postdoctoral Research Fellow}
\footnotetext[5]{Present address: Naval Research Laboratory,
Code 7213, 4555 Overlook Ave Sw, Washington DC 20375}
\footnotetext[6]{Department of Astronomy, University of Florida,
211 Bryant Space Science Center, P.O.~Box~112055, Gainesville, FL 32611;
guzman@astro.ufl.edu}
\footnotetext[7]{Institut d'Estudis Espacials De Catalunya/CSIC, Gran
Capit\`a 2-4, E-08034 Barcelona, Spain; fjc@ieec.fcr.es}
\footnotetext[8]{Apache Point Observatory, P. O. Box 59, Sunspot, NM 88349}

\begin{abstract}
We present the results of a single-dish beam-matched survey of the three
lowest rotational transitions of CO in a sample of 20 local 
(D $\lesssim$ 70 Mpc) Luminous Compact Blue Galaxies (LCBGs).  
These $\sim$L$^\star$, blue, high surface brightness, starbursting
galaxies were selected with the same criteria used to define LCBGs
at higher redshifts.  
Our detection rate was 70\%, with those galaxies having 
L$_B$~$<$~7~$\times$~10$^9$~L$_\odot$ not detected.
We find the H$_2$ masses of local LCBGs range from
$6.6\times~10^6$ to $2.7\times~10^9~M_\odot$, assuming a Galactic
CO-to-H$_2$ conversion factor.  Combining these results with our earlier 
H~{\small{I}}
survey of the same sample, we find that  
the ratio of molecular
to atomic gas mass is low, typically $5-10\%$.  
Using a Large Velocity Gradient model, we find that the average gas
conditions of the entire ISM in local LCBGs are similar
to those found in the centers of star forming regions in our Galaxy,
and nuclear regions of other galaxies.
Star formation rates, determined from IRAS fluxes, are a few
$M_\odot$~year$^{-1}$, much higher per unit dynamical mass than
normal spirals. If this rate remains constant, the molecular hydrogen
depletion time scales are short, $\sim 10-200$~Myr.
\end{abstract}


\section{Introduction}
Luminous Compact Blue Galaxies (LCBGs) are a morphologically and
spectroscopically diverse class of $\sim$L$^\star$,
blue, high surface brightness, high metallicity, vigorously 
starbursting galaxies with an underlying older stellar population
\citep[e.g.][]{1997ApJ...489..559G}.
While they have small optical diameters 
($\sim$few kpc) they are more luminous and metal-rich
than the widely studied local Blue Compact Dwarf Galaxies 
\citep[e.g.][]{1981ApJ...247..823T, 1994AJ....107..971T}.

LCBGs are common at intermediate redshifts
(0.4~$\lesssim$~z~$\lesssim$~0.7), but rare locally 
\citep{1999ApJ...518L..83M}.  They undergo
dramatic evolution, with their number density and star formation
rate density decreasing by at least a factor of ten from z~$\sim$~1 to
today \citep{1994AJ....108..437M, 1997ApJ...489..559G}.  
We have undertaken a survey of 
the neutral interstellar medium in 20 nearby (D $\lesssim$ 70 Mpc) LCBGs.
The galaxies are listed in Table \ref{t:basic_prop} and were primarily chosen
from the Sloan Digital Sky Survey
\citep[SDSS;][]{2000AJ....120.1579Y, 2004AJ....128..502A}
using a similar selection criteria to \cite{2004ApJ...617.1004W}
to be local analogs to the numerous LCBGs studied at intermediate redshifts.

In \citet[hereafter Paper I]{2004ApJ...615..689G},
we reported the results of a single-dish survey of 
the H~{\small{I}} content of these galaxies and showed that
LCBGs are gas-rich with velocity dispersions and optical sizes that
are consistent with low-mass galaxies, despite their high luminosities.
Many have disturbed morphologies and nearby companions at similar
velocities. We direct the reader to Paper I for complete information
on the definition of LCBGs, sample selection, optical photometric properties,
and results from the H~{\small{I}} survey.

The H~{\small{I}} observations do not tell us, however, about the properties of
the ISM that is actively forming stars. For this, it is necessary to
study the molecular gas.
In this work, Paper II, we report the results of a survey of the three
lowest rotational transitions of CO of the same sample of 20 nearby LCBGs.
CO is the most abundant,
readily observable molecule in star forming clouds. The intensity of 
its lowest transition, 1--0, is empirically found to be a direct
measure of the column density of molecular gas
\citep{1978ApJS...37..407D}, and the ratio of the different
transitions provides information on the temperature and density
of the gas \citep{1974ApJ...189..441G}.
The CO observations are described in \S\ref{s:obs},
and the derived molecular masses, densities and temperatures
are presented in \S\ref{s:detect}-\ref{s:multitrans}.
The rate at which the molecular gas forms stars is proportional
to the far-infrared (IRAS) luminosity and this is discussed,
along with gas depletion timescales, in \S\ref{s:tau}.
Our conclusions are listed in \S\ref{s:conclusions}.
We assume H$_0$~=~70~km~s$^{-1}$~Mpc$^{-1}$ throughout. 

\section{Observations}\label{s:obs}
We observed the sample of 20 nearby LCBGs in the CO(2--1) transition
at the JCMT 15~m telescope and followed up clear or marginal detections
with observations of CO(1--0) at the IRAM 30~m.
Most of these were also observed in CO(3--2) at the CSO and HHSMT 10~m
telescopes. The variety of telescopes was chosen to provide a uniform beam
size, $\simeq 21''$, in each transition. The observational setup for each
telescope is described below:

\subsection{JCMT}
The James Clerk Maxwell Telescope\footnote{
The JCMT is operated by the Joint Astronomy Centre in Hilo, Hawai`i 
on behalf of the parent organizations Particle Physics and Astronomy
Research Council in the United Kingdom, the National Research Council
of Canada and The Netherlands Organization for Scientific Research.} (JCMT)
was used to observe the CO(2--1) transition at 230~GHz
in queue mode observations between August 2002 and January 2003.
The heterodyne A3 receiver was used with the Digital Autocorrelating 
Spectrometer (DAS), giving a bandwidth of 920~MHz ($\sim$1200~km s$^{-1}$).
System temperatures ranged from 400 to 600~K
and sources were typically observed until the rms noise per
10~km~s$^{-1}$ channel was 10~mK, or the peak signal-to-noise was
greater than five.

\subsection{IRAM}
The Institut de Radioastronomie Millim\'etrique (IRAM) 30~m Telescope was 
used to observe the CO(1--0) transition at 115~GHz between August 2001 and
October 2003 using the A100 and B100 heterodyne receivers.
The 1~MHz filterbank was split in half giving a bandwidth of 512~MHz 
($\sim$1300~km~s$^{-1}$) in 2003 October, and was split in four 
giving a bandwidth of 
256 MHz ($\sim$670~km~s$^{-1}$) in 2002 May and 2001 August.  
System temperatures averaged 500~K
and sources were typically observed until the noise was
5~mK per 10~km~s$^{-1}$ channel,
or a peak signal-to-noise greater than five was reached.

The CO(2--1) transition was observed simultaneously with the
A230 and B230 receivers. The autocorrelator backend was configured
to a bandwidth of 320~MHz ($\sim$415~km~s$^{-1}$) per receiver.
The resolution of these data is $11''$.

\subsection{HHSMT}
Observations of CO(3--2) were performed with the Heinrich Hertz
Sub-Millimeter Telescope (HHSMT) in February 2002.
Mrk~297, Mrk~325 and SDSSJ0934+0014 were 
observed using the double sideband dual channel SIS-345 GHz
heterodyne receiver B30/B40, each with a 1~GHz 
($\sim$870~km~s$^{-1}$) Acousto-Optical-Spectrometer (AOS) backend.
System temperatures ranged from 700 to 800~K.

\subsection{CSO}
Two observing runs in March and July 2003 with the Caltech Submillimeter
Observatory (CSO) completed the CO(3--2) observations listed in
Table \ref{t:co_line_all}.
The 345~GHz heterodyne receiver and 1.5~GHz ($\sim$1300~km~s$^{-1}$)
AOS backend were used.
System temperatures averaged 700~K
and sources were typically observed until the rms noise
was $\sim$10 mK per 10 km s$^{-1}$ channel,
or the peak signal-to-noise was greater than five.

For each telescope, the observations were performed in beam-switching mode
with a throw of $240''$ and rate between 0.5 and 1~Hz.
System temperature calibrations were taken roughly every 20 minutes,
as well as each time a new source was observed. Pointing and main beam
calibration scans were done on planets or bright carbon stars
at least once every three hours. The focus was checked several
times each night, particularly near sunrise and sunset.

\subsection{Reduction and Calibration}
All data were reduced in a similar manner, using the
Continuum and Line Analysis Single-dish
Software (CLASS) for the IRAM, HHT and CSO data, while the JCMT 
data were reduced using SPECX.
A first or second order baseline was removed from each scan, avoiding the 
region of line emission. If no emission line was obvious in the scan,
a region of at least $\pm$50~km~s$^{-1}$
from the optical velocity was avoided. The individual scans were then
combined, and a first order baseline removed.  
Spectra were converted to the heliocentric velocity scale,
using the optical convention, and smoothed to a common resolution of
$\sim$10~km~s$^{-1}$, using a boxcar smoothing routine.
The IRAM data for Mrk~297 are from \cite{1993A&A...273....6S}
and were digitized from plots kindly provided by Leslie Sage.

The data were calibrated from raw intensity to the scale of corrected 
antenna temperature ($T_{\rm A}^*$)
at each telescope using the conventional ``chopper wheel method'' where loads
are measured at ambient and cold temperatures.  
The spectra were then calibrated from this telescope dependent quantity
to main beam temperature (T$_{mb}$).  
Both IRAM and JCMT maintain web-based databases of main beam efficiencies,
$\eta_{mb}$, which we used for calibration. For the IRAM data, 
$\eta_{mb}$~=~0.79 for the 115~GHz data, and $\eta_{mb}$~=~0.57 for
the 230~GHz data. The JCMT main beam efficiencies ranged from
$\eta_{mb}=0.55$ to 0.71 depending on the month of observation.
The HHT spectra were calibrated by comparing observations of
IRC+10216 with the spectral line survey of
\cite{1994ApJS...94..147G}, as well as by observations of Mars.  
We used $\eta_{mb}$~=~0.59 for
receiver B30 and $\eta_{mb}$ = 0.48 for receiver B40.  
We also calibrated the CSO data using observations of IRC+10216, finding
$\eta_{mb}$~=~0.60.

\section{Results}\label{s:results}

\subsection{Line measurements and detectability}\label{s:detect}
\begin{figure}
\centerline{
\includegraphics [width=7in] {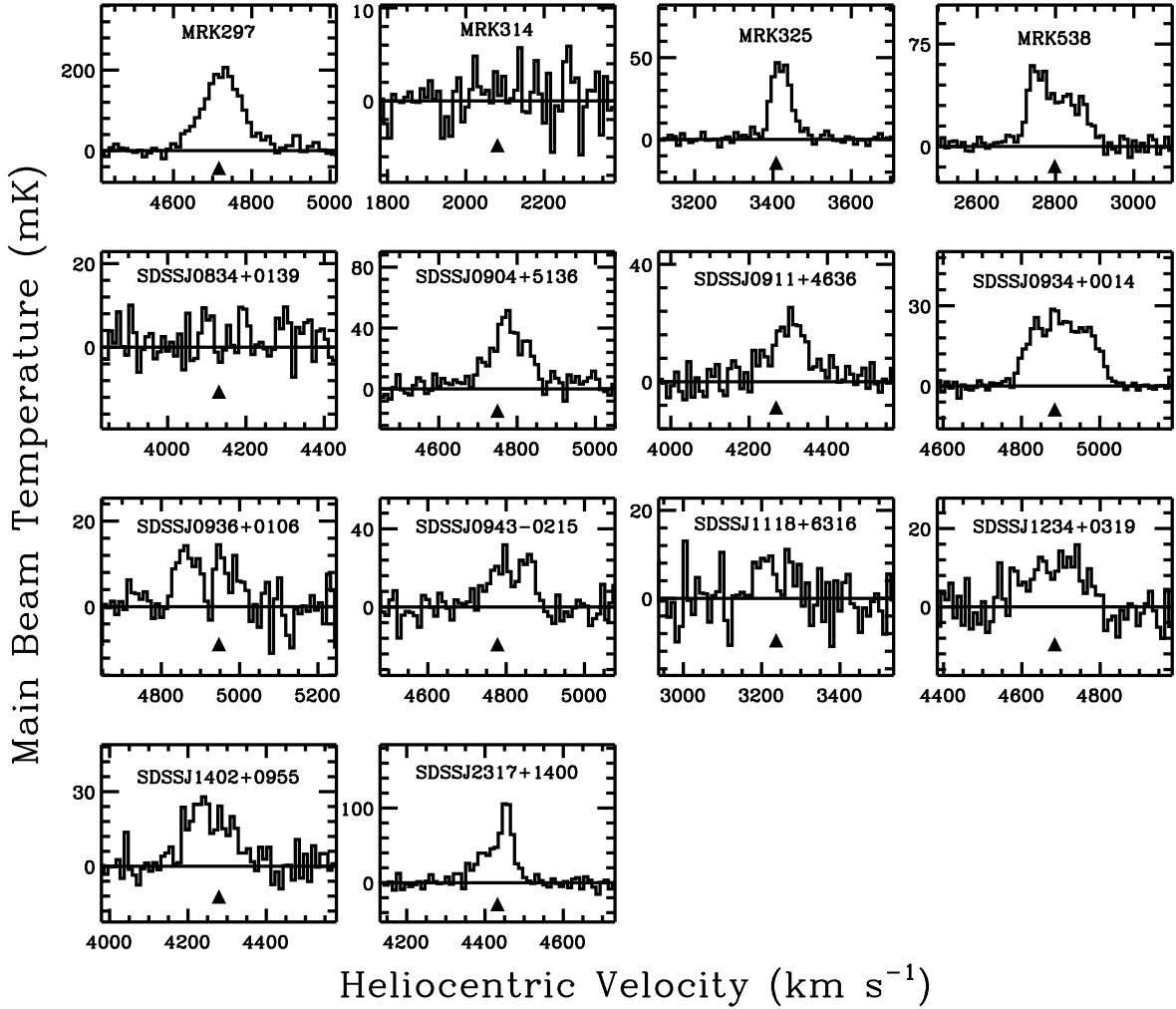}
}
\caption{
IRAM 30 m CO(J=1$-$0) spectra of the sample of nearby
LCBGs. The vertical scales are main beam temperatures, in mK; the 
horizontal scales are heliocentric velocities, in km s$^{-1}$.
The spectra have been smoothed to a resolution of $\sim$10
km s$^{-1}$ and only
the central 600 km s$^{-1}$ are shown. The triangles indicate
the recessional velocities calculated from SDSS redshifts for
the SDSS galaxies; velocities from NED redshifts are shown for the 
non-SDSS galaxies.
\label{f:IRAM10all}}
\end{figure}

\begin{figure}
\centerline{
\includegraphics [width=7in] {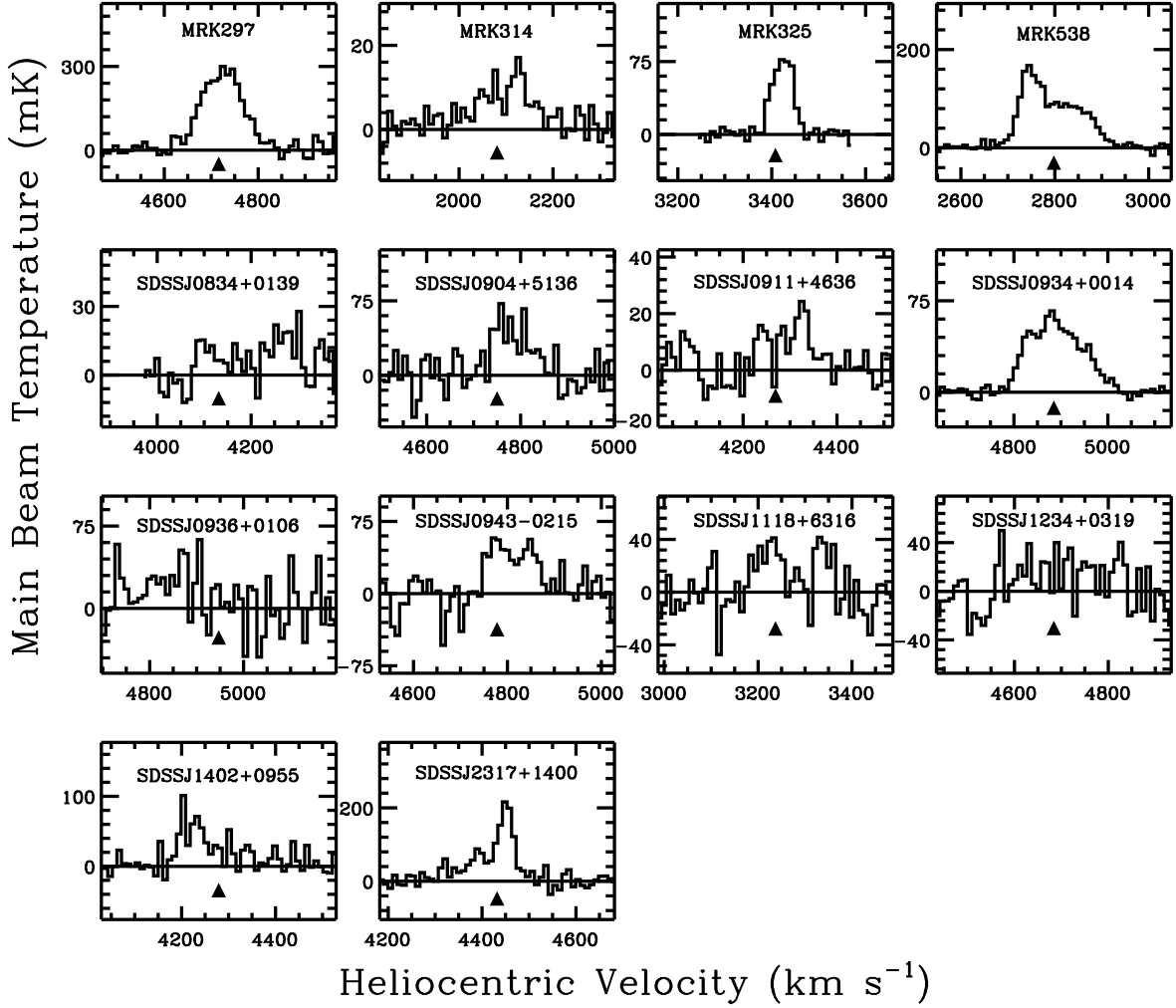}
}
\caption{
The central 500 km s$^{-1}$ of the
IRAM 30 m CO(J=2$-$1) spectra of the sample of nearby
LCBGs. The scales and labels are as in Figure
\ref{f:IRAM10all}.  Note that these observations
have a beam size half that of all others, 11$^\prime$$^\prime$.
\label{f:IRAM21all}}
\end{figure}

\begin{figure}
\centerline{
\includegraphics [width=7in]{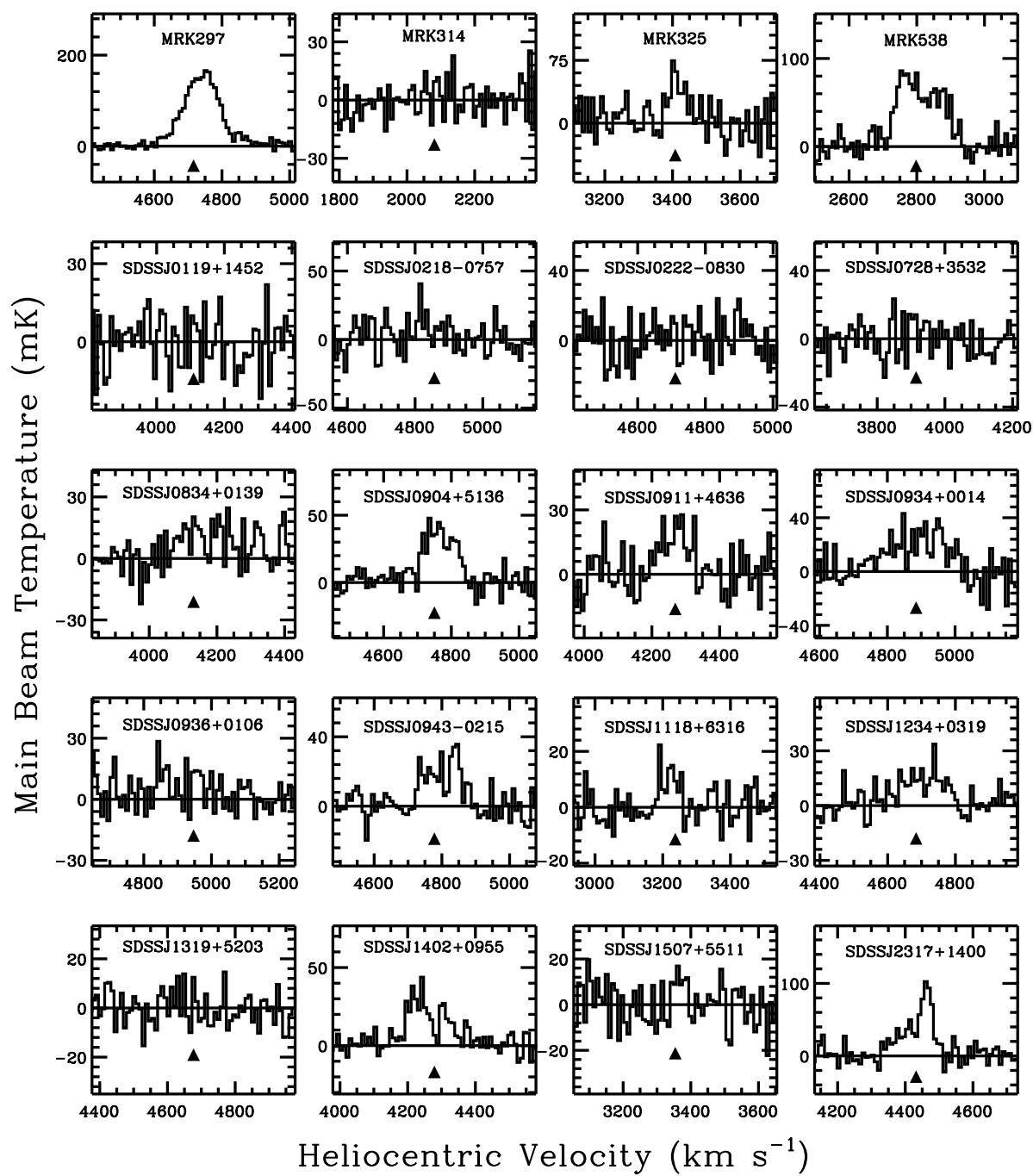}
}
\caption{
The central 600 km s$^{-1}$ of the
JCMT CO(J=2$-$1) spectra of the sample of nearby
LCBGs. The scales and labels are as in Figure
\ref{f:IRAM10all}.
\label{f:JCMT21all}}
\end{figure}

\begin{figure}
\centerline{
\includegraphics [width=7in]{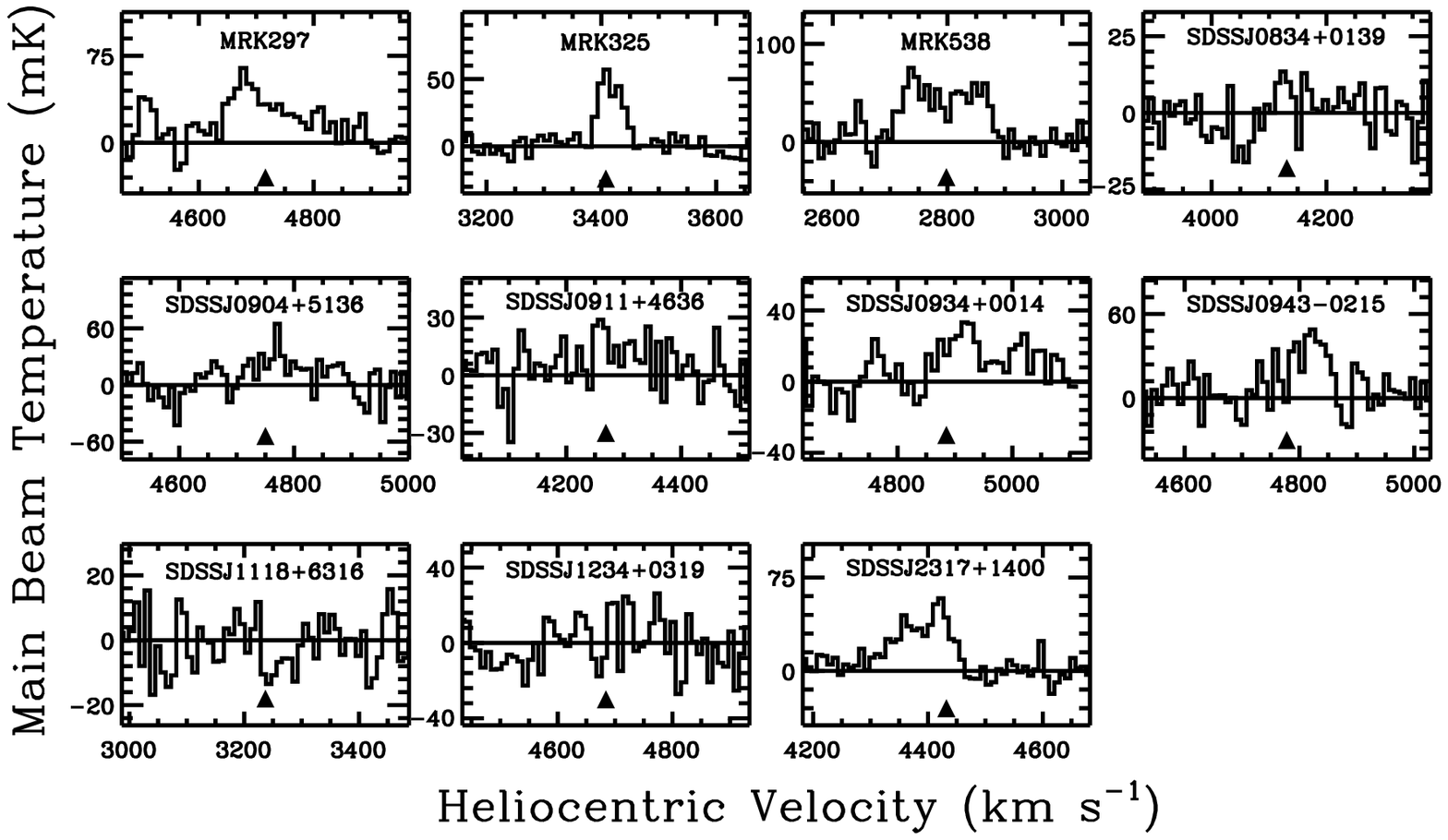}
}
\caption{
The central 500 km s$^{-1}$ of the
HHT and CSO CO(J=3$-$2) spectra of the sample of nearby
LCBGs. The scales and labels are as in Figure
\ref{f:IRAM10all}.
\label{f:CO32all}}
\end{figure}

The CO spectra are shown in Figures~\ref{f:IRAM10all}, \ref{f:IRAM21all},
\ref{f:JCMT21all}, and \ref{f:CO32all}.  Each galaxy's
velocity, calculated from the optical redshift, is indicated with a triangle.
The galaxies exhibit a variety of spectra, including Gaussian, double-horned 
and asymmetric profiles.
Line measurements, calculated by taking moments of the spectra,
are listed in Table \ref{t:co_line_all}.
We list the mean (heliocentric) velocity, $V_\odot$,
and dispersion, $\sigma$, from the CO(1$-$0) observations, except where noted.
In general, the central velocities and line widths agree well from 
transition to transition but the higher CO transitions
tend to have slightly smaller line widths. This may be due to the central 
concentration of the warmer, denser gas.
The CO velocity dispersions were always found to be smaller than the
H~{\small{I}} velocity dispersions calculated in the same manner (Paper I).

For those spectra with no discernible 
emission line, we calculated 3~$\sigma$ upper limits to the 
integrated intensity, $I_{CO}$, over the average
number of channels containing CO emission for that transition and 
beam size combination.  Blank entries in Table \ref{t:co_line_all} 
indicate the source was not observed with that transition and beam size
combination.
The uncertainties on $I_{CO}$ were calculated from the measured noise in the 
channels outside of the emission line region.
The uncertainties on the recessional velocities and line widths were 
estimated by creating 1000 simulated spectra from each observed spectrum, 
and measuring the
dispersion of the calculated recessional velocities and line widths.

Six galaxies, SDSSJ0119+1452, SDSSJ0218$-$0757, SDSSJ0222$-$0830, 
SDSSJ0728+3532, SDSSJ0834+0139, and SDSSJ1319+5203
were not detected in CO above the 3~$\sigma$ level.
We do not expect that beam dilution is a serious problem because
the typical optical diameter, $D_{25}$ (Table \ref{t:basic_prop}), 
is roughly twice the uniform $21''$ resolution of the data.
\cite{1995ApJS...98..219Y} found that CO extends over half the optical disk, 
on average, in a sample of over 300 nearby galaxies of varied Hubble types and
environments and therefore we expect the telescope beam sizes enclose most
of the CO emission for most of our sample. We did find that the detection
rate was lower for the smallest galaxies, those with $R_{25}\lesssim 21''$.
Regardless of size, all galaxies with luminosities,
$L_B>7\times 10^9~L_\odot$ ($M_B<-19.1$) were detected.

\subsection{Molecular Masses}\label{s:mh2}
We converted the CO integrated intensities to molecular hydrogen
masses using a Galactic conversion factor of
$1.8\times 10^{20}$~cm$^{-2}$~K$^{-1}$~km$^{-1}$~s
\citep{2001ApJ...547..792D}.  This ``X'' factor, is only well calibrated
for the Milky Way and Local Group.  It depends on the
physical conditions of the gas (e.g. metallicity, density, temperature)
which are difficult to determine and may differ greatly
from those in our own Galaxy \citep{1998AJ....116.2746T, 1992A&A...265...19S}.  
There is considerable debate over how X depends on galaxy properties,
particularly in low metallicity and/or starbursting galaxies 
\citep[e.g.,][]{1995ApJ...448L..97W, 2003ApJ...588..771Y}
and also, interestingly, resolution \citep{2003ApJ...599..258R}.
The consensus from the latest studies is that X is not as
sensitive to metallicity as older lower resolution measurements
had suggested.

%It can also be argued that CO emission is so optically thick that metallicity
%is probably not the dominant uncertainty in measuring the molecular mass
%content of a galaxy. The magnitude of the effect can be gauged in our own
%Galaxy by considering $^{13}$CO, which is $\sim$60 times less abundant 
%than $^{12}$CO \citep[e.g.,][]{1994ARA&A..32..191W}
%but only $\sim$5~$-$~10 times less intense in the 1--0 transition
%for Giant Molecular Clouds in the Galaxy (e.g. \cite{1988ApJ...332..432P}.) 
We also expect local LCBGs to have near-solar metallicities based on 
observations of intermediate redshift LCBGs
\citep{1996ApJ...460L...5G, 2000ApJ...545..712K},
and the four local LCBGs for which there are metallicity estimates in the
literature \citep{1997AJ....113..162C, 1986PASP...98....5H}.
Based on this and the lack of any clear dependence of X on
metallicity, we adopt the Galactic value for converting our
measured CO intensities to molecular masses.

The empirical relation for the mass of H$_2$ is calibrated for the 
CO(1--0) transition.
We have observed 14 galaxies in CO(1--0), and detected 12 of those.  
The two non-detections, Mrk 314 and SDSSJ0834+0139, are treated differently:
Mrk~314 was detected in CO(2--1) with the larger collecting area and smaller
beam, $11''$, of the IRAM 30~m and we use the integrated intensity
of that spectrum to convert to a column density and then mass.
SDSSJ0834+0139 was undetected even at higher resolution and we use the
3~$\sigma$ upper limit from the CO(1--0) observations.
For the galaxies not observed in CO(1--0) at IRAM, we used the
CO(2--1)~$\theta=21''$ observations, either $3~\sigma$ upper limits
or detections from the JCMT data. It is reasonable to estimate the H$_2$
mass using CO(2--1) instead of CO(1--0) as we find the ratio of these
integrated intensities is 1.1 $\pm$ 0.21.

The H$_2$ masses of local LCBGs 
are listed in Table \ref{t:co_line_all} and range from 
$6.6\times 10^6$ to $2.7\times 10^9~M_\odot$, with a median
of $1.8\times 10^8~M_\odot$ (including upper limits).
Molecular hydrogen masses range from less than 
$1\times 10^6$ to $5\times 10^{10}~M_\odot$ across
the range of galaxy types \citep{1991ARA&A..29..581Y},
so LCBGs are not extreme in either direction.
The most massive LCBGs have molecular masses comparable to
the Milky Way but the average is an order of magnitude smaller.
We also note that the H$_2$ masses are all less than the upper limits
in $z\sim 0.5$ LCBGs obtained by \cite{1998A&A...330...63W}.

\subsection{Comparison with HI observations}\label{s:hi}
\begin{figure}
\centerline{
\includegraphics [width = 7in] {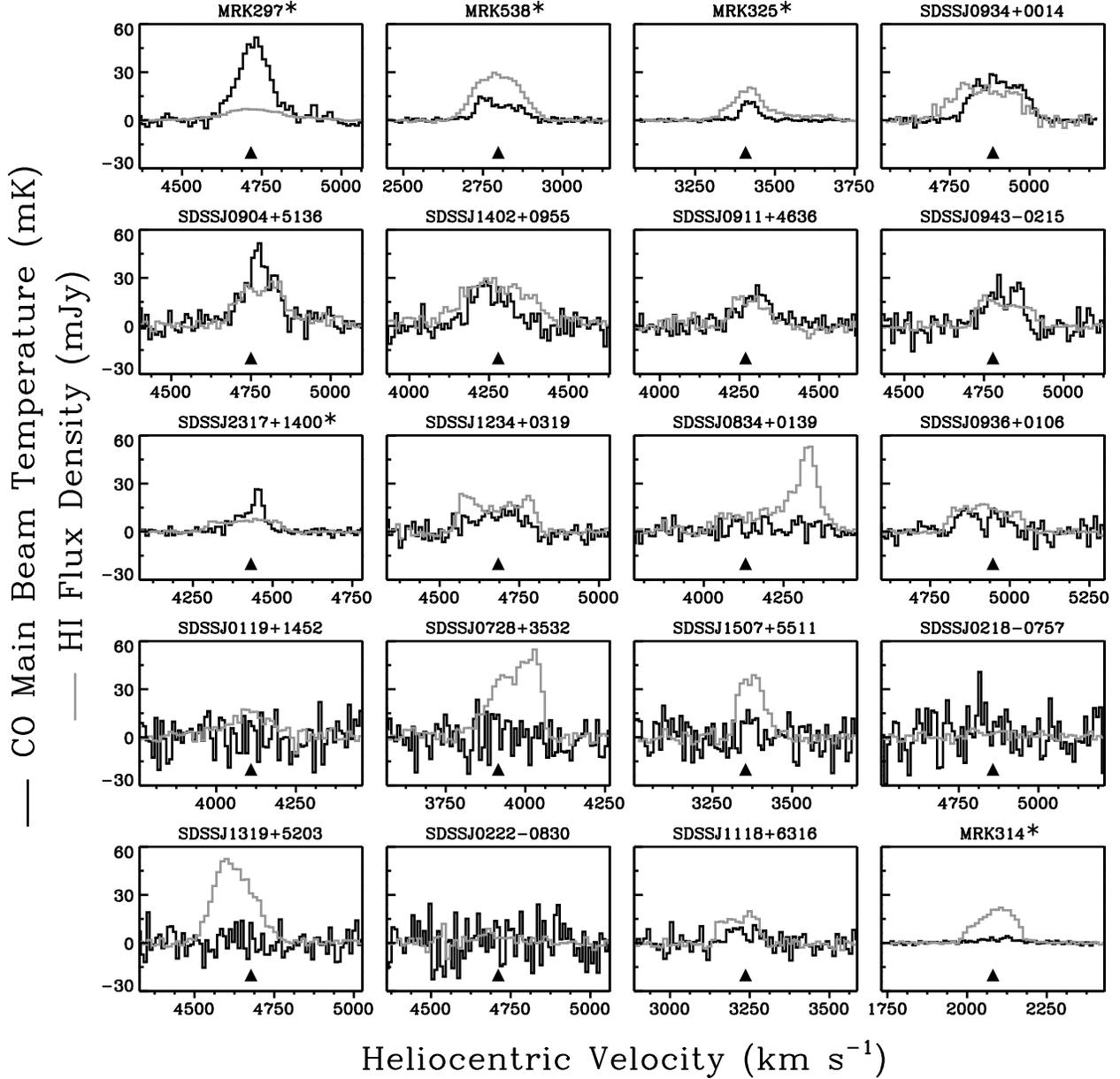}
}
\caption{Comparison of CO and HI spectra for the 20 galaxies in
our sample, ordered from brightest to faintest absolute blue
magnitude. The CO spectra are shown as the darker line and either
IRAM or JCMT data (see text). The gray line shows the HI spectra
from Paper I and the triangle indicates
the velocity of each galaxy measured from optical emission lines.
All galaxies are plotted on the same temperature/flux density scale
except for five galaxies, indicated by asterisks, where the spectra
have been scaled down by a factor of four.
\label{f:CO_HI}}
\end{figure}

It is instructive to compare these observations of the molecular
ISM with the observations of the neutral atomic ISM in Paper I.
CO and HI spectra are plotted for each of the 20 galaxies in our
sample in Figure~\ref{f:CO_HI}. The CO spectra are generally the
1--0 transition from the IRAM 30~m observations ($21''$ resolution)
except for Mrk 314, which was only detected in the 2--1 transition
(at $11''$ resolution), and the six galaxies,
SDSSJ0119, 0218, 0222, 0728, 1319 and 1507 which were not
observed with IRAM, and where we show the JCMT CO(2--1)
spectra ($21''$ resolution) instead.
The HI spectra were taken with the Green Bank Telescope
(GBT; $9'$ resolution) and are discussed in more detail in Paper I.
For ease of comparison, the galaxies are ordered from
highest blue luminosity to lowest and are shown over the
same velocity width, 650~km~s$^{-1}$. The temperature/flux
density scale is also the same for all but five galaxies,
indicated by asterisks in the title to each panel, where the
spectra are scaled down by a factor of four.

\begin{figure}
\centerline{
\includegraphics [width = 7in] {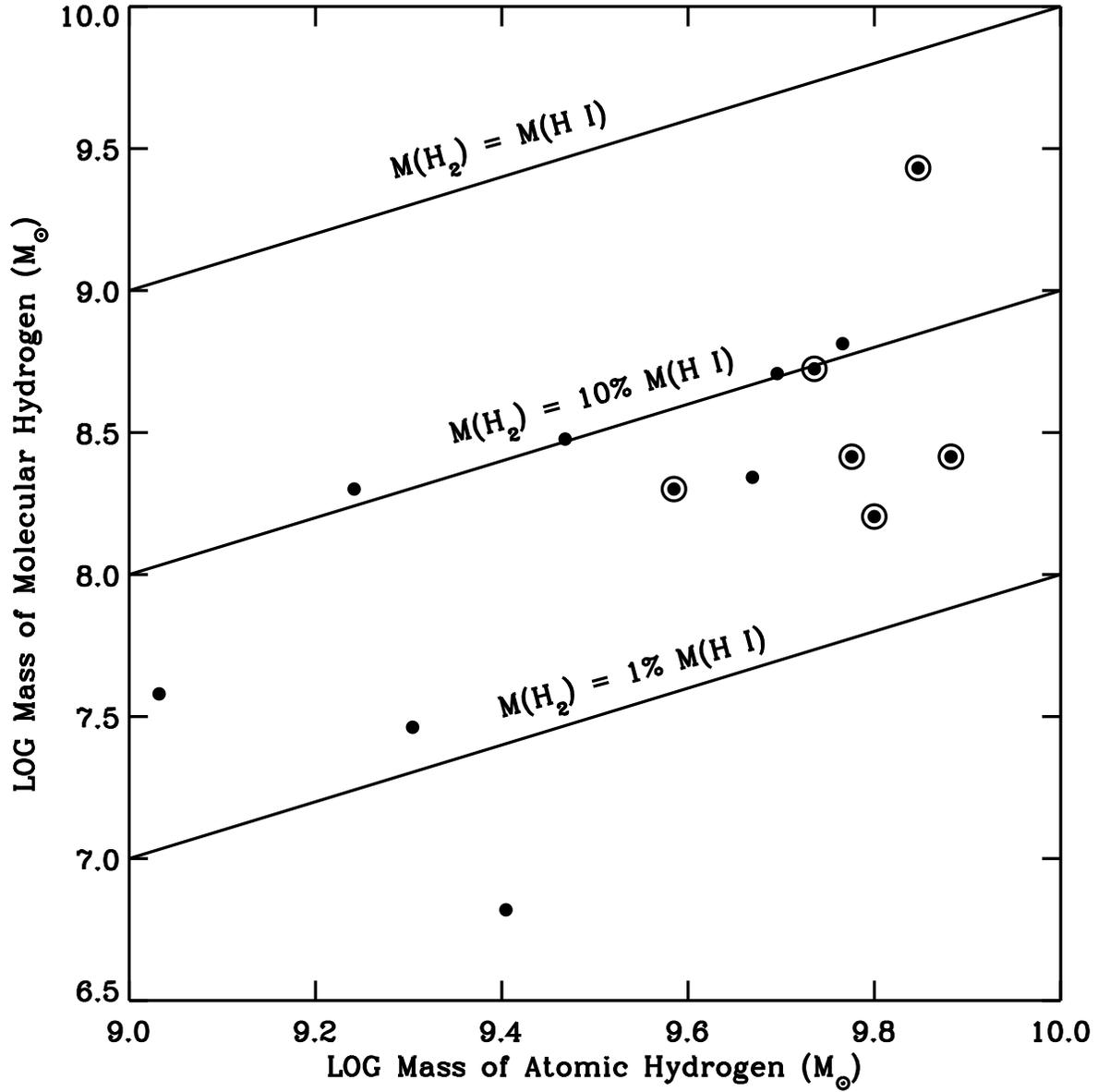}
}
\caption{Molecular versus atomic hydrogen mass for the sample of
local LCBGs. Galaxies with companions within the $9'$ GBT beam, for
which the H~{\small{I}} mass may be overestimated, are circled.
\label{f:MH2_MHI}}
\end{figure}

The HI spectra generally have a much higher signal-to-noise
ratio than the CO but, nevertheless, it is clear that the
CO-to-HI ratio varies greatly across our sample.
In Figure \ref{f:MH2_MHI} we plot the mass of molecular versus atomic
hydrogen for the 14 LCBGs detected in CO.
The ratio of molecular to atomic gas masses is low, typically $\lesssim 10\%$.
\cite{1989ApJ...347L..55Y} showed that the mean molecular to atomic mass
fraction varied systematically along the Hubble sequence from
from $2.6\pm 1.2$ for type S0/Sa to $0.12\pm 0.064$ for type Sd/Sm
(for $X = 1.8 \times 10^{20}$~cm$^{-2}$ K$^{-1}$ km$^{-1}$ s.)
LCBGs have a mean $M(H_2)$ to $M(H~{\small{I}})$ ratio 
of $0.05\pm 0.002$, even smaller than the median for late-type local 
spirals. Within our LCBG sample, however, we were unable to find
a correlation between a galaxy's morphology in the SDSS optical
images and its molecular-to-atomic mass ratio.

\begin{figure}
\centerline{
\includegraphics [width=7in] {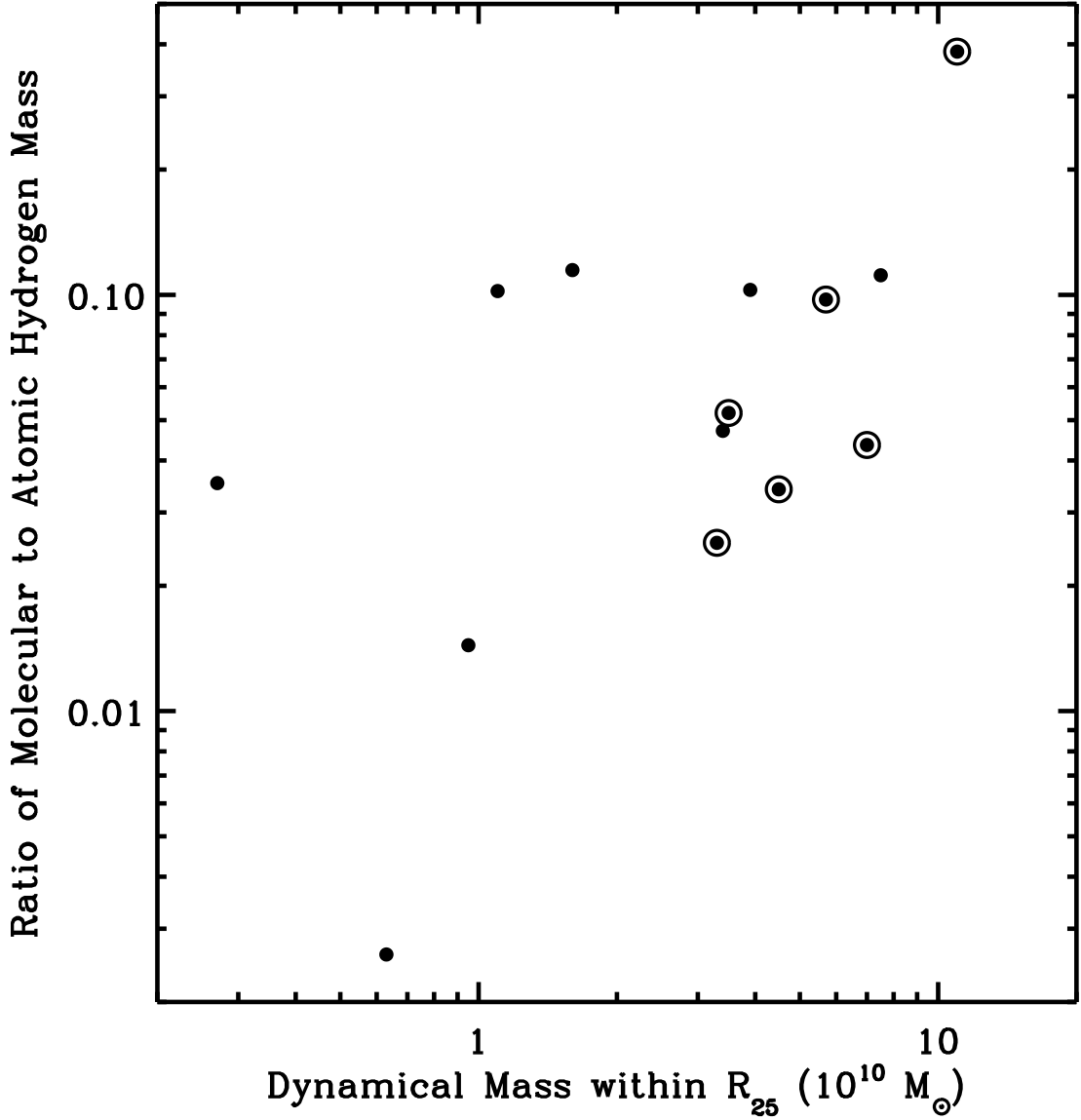}
}
\caption{The dynamical mass (within $R_{25}$) is plotted versus the
molecular to atomic mass fraction in local LCBGs. Circled galaxies
have companions within the $9'$ GBT beam and may have slightly
overestimated H~{\small{I}} masses. Nevertheless, there is a slight overall 
trend
for the fraction of molecular hydrogen to increase with increasing
dynamical mass.
\label{f:ratio_mdyn}}
\end{figure}

\cite{1989ApJ...347L..55Y} found that it is the phase,
not the amount, of neutral gas that changes with Hubble type.
They argue that the bulk of the decrease in the ratio of $M(H_2)$ to 
$M(H~{\small{I}})$ with Hubble type is from different
surface densities of the gas in the inner disks of early versus
late-type spirals. Blitz \& Rosolowsky (2004) hypothesize that the
hydrostatic pressure, set by the {\it stellar} surface density,
determines how much of the ISM is molecular or atomic at each
radius. There is some support for these ideas in the
weak correlation between the molecular to atomic mass
fraction and dynamical mass, shown in Figure \ref{f:ratio_mdyn}.

\subsection{H$_2$ densities and temperatures}\label{s:multitrans}
Comparing the integrated intensities of different CO transitions observed
with the same beam size allows us to probe the excitation conditions in the 
molecular gas.
As we are dealing with single-dish unresolved observations, our findings
are for the \emph{average} conditions in the molecular gas in the galaxies.  
We list the observed line ratios in Table \ref{t:line_ratios}, 
defined as
\[
\mathcal{R}_{ij} =
       \frac {I_{CO}~(i\rightarrow i-1)} {I_{CO}~(j\rightarrow j-1)},
\]
for $(i,j)=(3,1), (3,2)$ and $(2,1)$.

The noise in the JCMT CO(2--1) spectrum of Mrk~325 was
quite high, leading to large uncertainties in the line ratios.  
While observing Mrk~325 with the IRAM 30~m, we did a five point map
of the galaxy, with observations spaced by 
$\pm 10''$ from the central galaxy
position.  Using a weighted sum, we created a 2--1 spectrum to mimic
a $21''$ beam.  We found that this spectrum of Mrk~325
matched the JCMT CO(2--1) $\theta = 21''$ spectrum, but
with significantly better signal-to-noise.  Therefore,
we utilized the values from the IRAM CO(2--1) observations of 
Mrk~325 in our line ratio analysis.

In the sample of local LCBGs, 
$\mathcal{R}_{21}$ ranges from 0.83 to 1.6,
with a mean of $1.1\pm 0.21$. 
Our observations are in good agreement with those of 
\cite{1995A&A...300..369A} who found
$\langle \mathcal{R}_{21}\rangle = 0.93\pm 0.22$
in a study of a variety of infrared-bright galaxies.  
Note, however, that their observations
only cover the central positions of the galaxies. 
Optically thick CO emission
arising from thermalized gas at $T>10$~K yields 
$\mathcal{R}_{21}\sim 1$
assuming a homogeneous source
characterized by a single temperature and density
\citep{2003ApJ...595..167B}.  
Ratios lower than unity can be caused by low temperature
or volume density or departures from homogeneity.
For example, $\langle\mathcal{R}_{21}\rangle = 0.5$
in the Taurus Dark Cloud 
with slightly higher values $\mathcal{R}_{21} \gtrsim 0.7$
found in Giant Molecular Clouds, such as Orion A
\citep{1996IAUS..170...39H, 1996ibms.conf.....K}.

We find $\mathcal{R}_{31}$ ranges from 0.24 to 1.2, 
with a mean of $0.72\pm 0.33$, in the local sample of LCBGs.  
\cite{2003ApJ...588..771Y} studied the central regions of 
60 infrared luminous galaxies and found 
$\mathcal{R}_{31}$ to range from 0.22 to 1.72, with a median of 0.66.
These findings are similar to a study by \cite{1999A&A...341..256M}
who found $\mathcal{R}_{31}$ to range from 0.1 to 1.6, with a median
of 0.63, in a sample of nearby spiral galaxies.
They note that, as many of their galaxies are extended with respect
to the beam, their findings may reflect conditions in the nuclear
region. \cite{2001AJ....121..740M} found $\mathcal{R}_{31}$ to range
from 0.37 to 1.1, with a median of $0.61 \pm 0.06$ in a sample of dwarf 
(low luminosity) starburst galaxies.
These median values are larger than the average over a Milky Way
giant molecular cloud ($\mathcal{R}_{31}=0.4$) but similar to
the denser clumps within it ($\mathcal{R}_{31}=0.6-1$)
\citep{1998ApJ...494..657W, 1993AIPC..278..311S}.
That is, the \emph{average} conditions of the entire molecular ISM
in our LCBGs are similar to the conditions found in the \emph{cores}
of star forming regions in our own Galaxy and the
\emph{centers} of other galaxies.

We can further constrain the physical conditions of the molecular gas
using a Large Velocity Gradient \citep[LVG;][]{1974ApJ...189..441G}
model for the eight galaxies detected in all three transitions of CO.
In the models, we utilized an inferred column density of CO, N$_{CO}$, 
assuming a CO abundance of 10$^{-4}$ relative to H$_2$
\citep{1999RvMP...71..173H}, and CO-to-H$_2$ conversion factor of
$1.8\times 10^{20}$~cm$^{-2}$~K$^{-1}$~km$^{-1}$~s (Table \ref{t:lvg}).  
We ran the models for a grid of kinetic temperatures, $T_K$, between
5 and 200~K, and volume densities, n(H$_2$), between
10 and $10^7$~cm$^{-3}$. We then derived model intensity ratios,
$\mathcal{R}_{21}, \mathcal{R}_{31}$ and $\mathcal{R}_{32}$
and searched for temperature and density combinations that matched
the corresponding three observed values.

\begin{figure}
\centerline{
\includegraphics [width=7in] {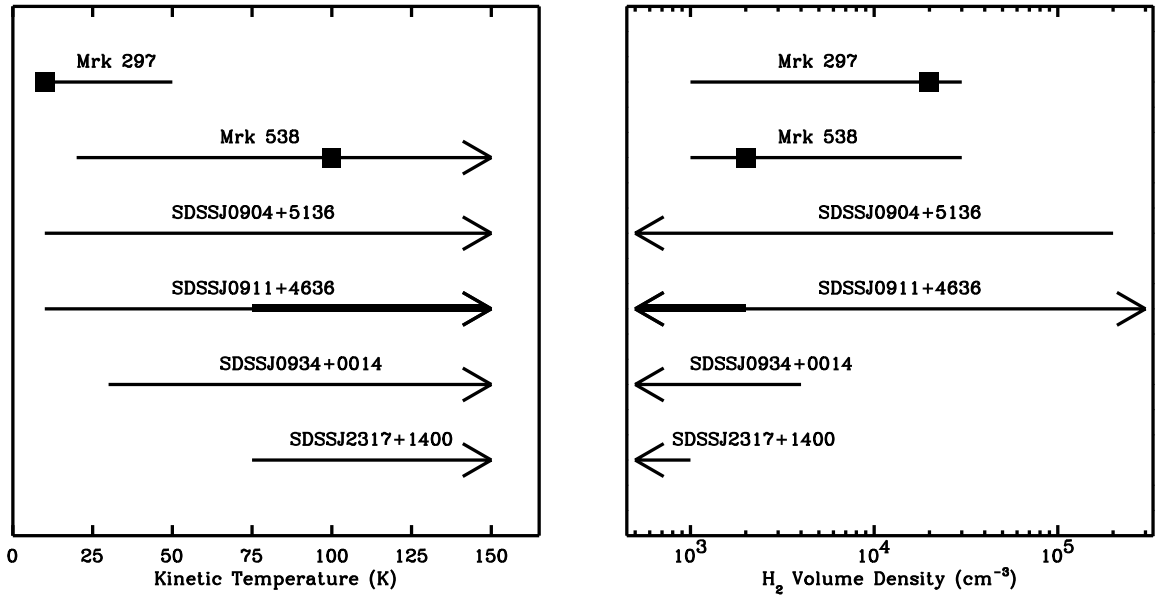}
}
\caption{
The ranges of possible kinetic temperatures (left)
and molecular hydrogen volume densities (right)
for local LCBGs as predicted by LVG modeling.
Model fits to within $\pm$3$\sigma$ of the observed
values are indicated by thin lines/arrows, while those
to within $\pm$1$\sigma$ are indicated by heavy lines
or dark rectangles (where only one solution was found
in the model grid).
\label{f:lvg_results}}
\end{figure}

The results of our LVG modeling of six LCBGs are shown in
Figure \ref{f:lvg_results}. We were able to fit all three ratios
within $\pm 1\sigma$ in three cases, Mrk~297, Mrk~538, and SDSSJ0911+4638.
A wide variety of temperature and density combinations were found,
however, ranging from $T=10$ to 150~K, and
$n({\rm H}_2)\lesssim 10^3$ to $2\times 10^4$~cm$^{-3}$.  
Our findings are similar to the study of dwarf starburst galaxies
by \cite{2001AJ....121..740M}.

For three other galaxies, SDSSJ0904+5136, SDSSJ0934+0014, and
SDSSJ2317+1400, we found fits that matched the ratios within $\pm 3\sigma$,
but these only provided lower temperature limits and upper density limits.
We were unable to find any combinations of temperature and volume 
density which worked simultaneously for all three line ratios for
the remaining two galaxies, Mrk 325 and SDSSJ0943$-$0215.
The temperatures and densities which match
$\mathcal{R}_{21}$ and $\mathcal{R}_{31}$ are inconsistent with
those for $\mathcal{R}_{32}$, even at very high
temperatures ($>$~200 K) and volume densities ($>$~10$^5$~cm$^{-3}$).
To produce a fit in these galaxies, we would need to increase 
N$_{CO}$ above the observed values
of $\sim$5~$\times$~10$^{16}$ cm$^{-2}$, by approximately an 
order of magnitude.
Note that our findings would change if we utilized different values of
$N_{CO}$.  For example, if
the X factor or the CO to H$_2$ abundance were higher, then the 
inferred $N_{CO}$ would
be larger, and lower temperatures and/or densities would be predicted.

\subsection{Star Formation Rates and Gas Depletion Time Scales}\label{s:tau}
We extracted InfraRed Astronomical Satellite (IRAS) fluxes at 
12, 25, 60 and 100~$\micron$ from the Faint Source Catalog v2.0 (FSC)
for all but one source, SDSSJ0222$-$0830, which was not cataloged.
Table~\ref{t:basic_prop} lists the infrared luminosities, estimated from 
the 60 and 100~$\micron$ fluxes based on the formulation of 
\cite{2002AJ....124.3135K}.  
Five galaxies in our sample have companions within $5'$
(Table~\ref{t:basic_prop}) and may be confused at 100~$\micron$.
The effect is likely to be minor: for example, Mrk~538 is resolved in the
IRAS High Resolution Image Restoration (HIRES) Atlas of \cite{hires}
but the 60 and 100 $\micron$ fluxes of the primary vary by less than
10\% from those in the FSC.

We used the infrared luminosities derived from the FSC fluxes
to calculate star formation rates (SFRs) following \cite{1994ApJ...435...22K}.
Note that these SFRs assume that young stars
dominate the radiation field in the ultraviolet to visible range 
\citep{2002AJ....124.3135K}.
This is a reasonable assumption for starbursting galaxies such as LCBGs.  
For example, \cite{2002MNRAS.332..283R} showed consistency between
SFRs derived from H$\alpha$ and IRAS fluxes for a sample of 31 nearby
star forming galaxies, such as H~{\small{II}} galaxies, starburst galaxies, 
and blue compact galaxies. As a check, we also derived SFRs
from the 1.4~GHz radio continuum for the 15 LCBGs cataloged in the 
National Radio Astronomy Observatory Very Large Array Sky Survey
\citep[NVSS;][]{1998AJ....115.1693C} using the prescription of
\cite{2003ApJ...586..794B} and found excellent agreement with the infrared
estimates. The correlation between radio continuum and far-infrared fluxes
is well established for all galaxy types \citep{1992ARA&A..30..575C}
and it is reassuring to see that LCBGs are not an exception to this relation.

The SFRs range from 0.4 to $14~M_\odot$~year$^{-1}$,
with a median of $1.5~M_\odot$ year$^{-1}$~(Table \ref{t:basic_prop})
for our sample of local LCBGs.
This is comparable to values measured for
intermediate redshift (0.2~$\lesssim$~z~$\lesssim$ 1) LCBGs 
from [O {\small{II}}] equivalent widths \citep{1997ApJ...489..559G}.
LCBGs have similar far infrared luminosities and SFRs to
more massive local, normal spiral galaxies \citep{1994ARA&A..32..115R}.
The \emph{specific} star formation rates (SSFR),
defined as the ratio of SFR to dynamical mass \citep{1997ApJ...489..559G},
is therefore considerably enhanced in LCBGs.
Local LCBGs have SSFRs, calculated using the dynamical mass 
within R$_e$(B) (Paper I),
ranging from 1.1 to $6.6\times 10^{-10}$~year$^{-1}$.
These SSFRs are much higher than those measured for quiescent 
early-type spirals, $<10^{-11}$~year$^{-1}$.
They are more similar to H~{\small{II}} galaxies, which have SSFRs ranging
from $1.3\times 10^{-10}$ to $4.1\times 10^{-9}$~year$^{-1}$,
and intermediate redshift (0.2~$\lesssim$~z~$\lesssim$~1) LCBGs,
with SSFRs between $2\times 10^{-11}$ to $1.3\times 10^{-9}$~year$^{-1}$
\citep{1997ApJ...489..559G}.

Determining how long a LCBG can continue to form stars at its current
rate provides a direct constraint on its evolutionary possibilities. 
We can estimate how long a starburst will last by calculating the gas 
depletion time scale, $\tau_{gas}=M_{gas}/SFR$.
In local LCBGs, the molecular gas depletion time scales, $\tau(H_2)$,
range from 10 to 200 Myr (Table \ref{t:co_line_all}).
The total gas depletion time may be as much as a factor of 20 times
longer because the molecular mass fraction is only $\sim 0.05$,
but this requires efficient conversion of atomic to molecular
gas. Even so, most LCBGs have gas depletion timescales that are
much shorter than the lookback time to their period of greatest number
density at $z\sim 1$ and we therefore expect most of them to fade
and redden as they evolve into other galaxy types.

\begin{figure}
\centerline{
\includegraphics [width=7in] {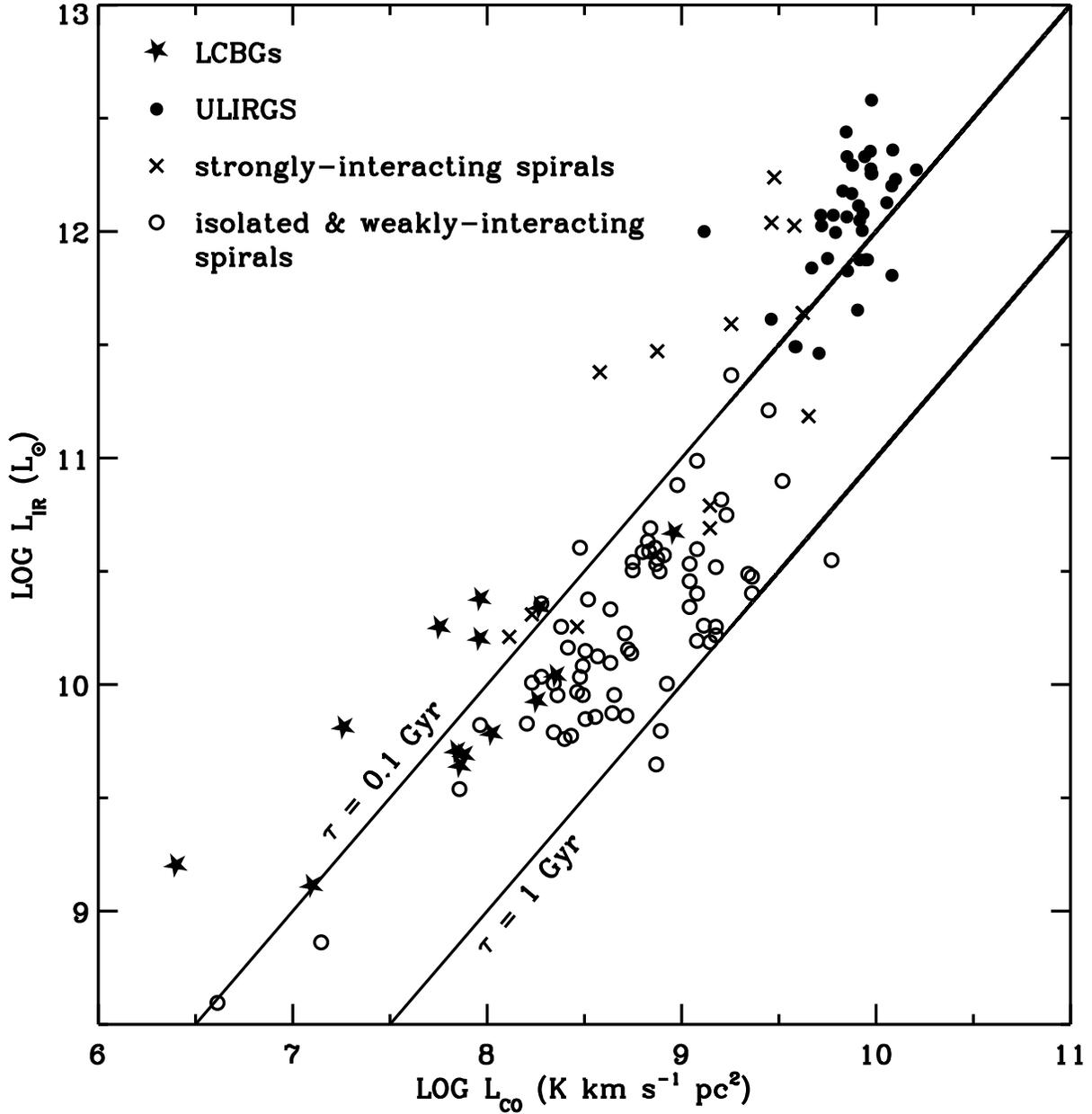}
}
\caption{
Infrared versus CO luminosities for local LCBGs (stars) and other 
galaxy types:  isolated or weakly-interacting local spirals (open circles), 
strongly-interacting local spirals ($\times$'s), and UltraLuminous InfraRed
Galaxies (ULIRGs, filled circles)
\citep{1997ApJ...478..144S, 1988ApJ...334..613S}.
Lines indicating gas depletion time scales of $10^8$ and $10^9$ years are
indicated.  LCBGs have short gas depletion time scales, similar to ULIRGs and
strongly-interacting local spirals.
\label{f:solomon}}
\end{figure}

An alternate way of representing gas depletion time scales is in terms
of infrared and CO luminosities
\citep[$L_{IR}$ and $L_{CO}$;][]{1997ApJ...478..144S}.
These are plotted against each other in Figure \ref{f:solomon}
for our sample of local LCBGs and compared to isolated galaxies,
interacting local spirals, and UltraLuminous InfraRed Galaxies (ULIRGs)
using values from
\cite{1997ApJ...478..144S} and \cite{1988ApJ...334..613S}.
The gas depletion time scale, proportional to $L_{CO}/L_{IR}$,
is shorter in LCBGs than in most isolated or
weakly-interacting spirals, but are similar to strongly-interacting
spirals and ULIRGs. LCBGs resemble scaled-down ULIRGs, in the sense
that they have high $L_{IR}$ to $L_{CO}$ ratios, but lower luminosities.
It is thought that the star formation efficiency is high, or the gas 
depletion time scale is short, in perturbed environments because
colliding gas flows can lead to high compression of gas 
\cite[e.g.][]{1993PASJ...45...43S}. Inspection of the SDSS optical
images, however, shows no obvious pattern between how disturbed
a galaxy appears and its location in this plot.

\begin{figure}
\centerline{
\includegraphics [width=7in] {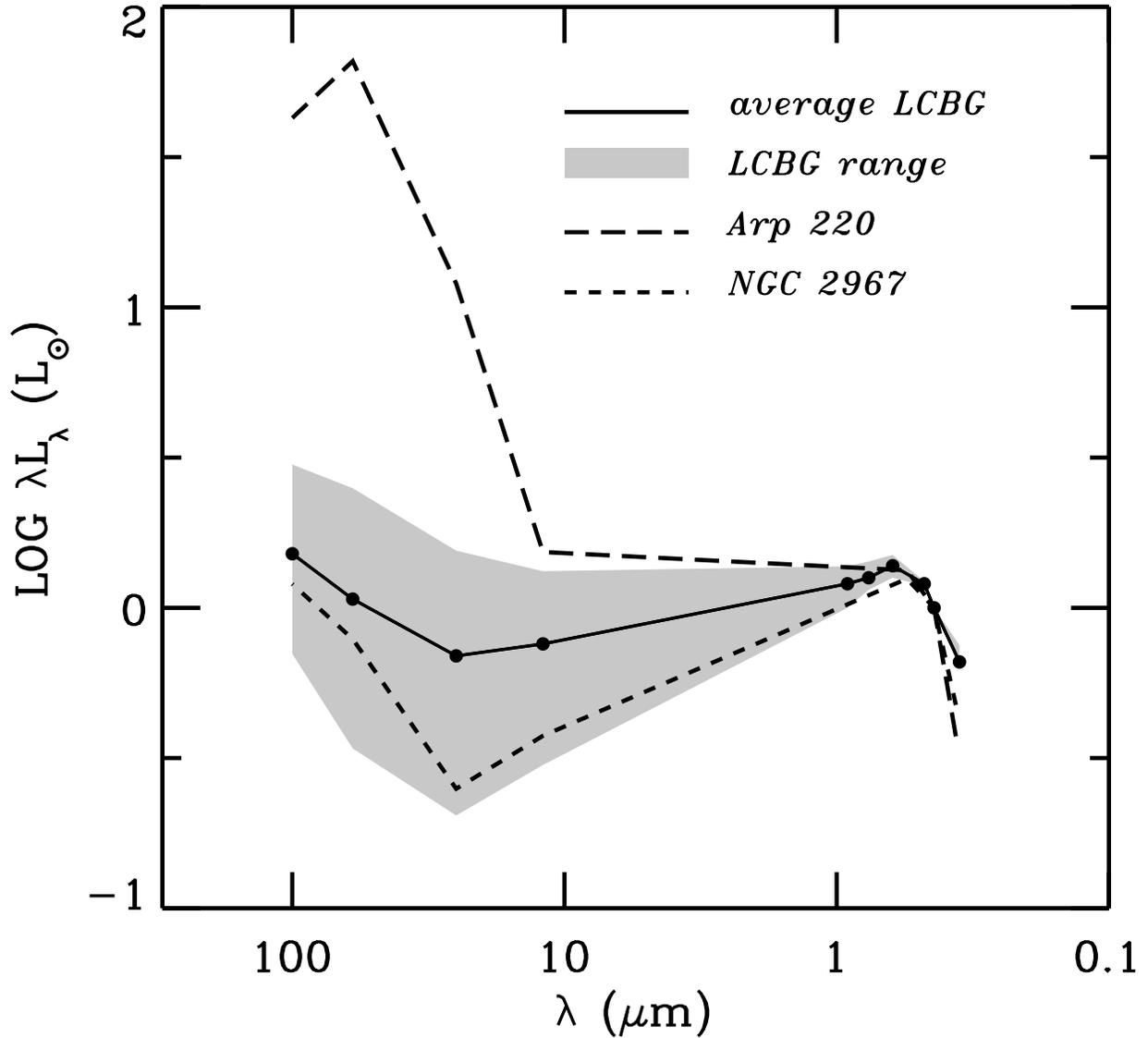}
}
\caption{Spectral energy distribution of LCBGs in comparison with
a normal spiral galaxy, NGC 2967, and a luminous infrared galaxy, Arp 220.
Optical fluxes were derived from the SDSS or NED where appropriate.
The far-infrared fluxes are from IRAS.
The average for the local LCBG sample is shown by the solid
line and the range by the gray area.
For comparison, each SED is normalized to the B-band value.
\label{f:SED}}
\end{figure}

Nevertheless, based on Figure~\ref{f:solomon}, can we consider LCBGs
to be simply miniature versions of (ultra-)luminous infrared and
interacting galaxies? If so, we would expect them to produce
most of their luminosity in the far-infrared. Spectral energy
distributions for the local LCBG sample are shown and compared
with a normal spiral, NGC 2967, and prototypical ULIRG, Arp 220,
in Figure~\ref{f:SED}. About equal proportions of the total
energy of an LCBG is produced in the far-infrared as in the optical.
This is more than for normal spirals but far less than the
dramatic infrared excesses found in strongly interacting
galaxies or ULIRGs \citep[e.g.][]{1996ARA&A..34..749S}.
This difference in the SEDs between ULIRGs and LCBGs cannot be
accounted for by lower metallicities or dust content in the latter
as only a small amount of dust is required to reprocess short
wavelength light to the far-infrared.
Thus the locations of LCBGs in Figure~\ref{f:solomon}, and the
inferred short gas depletion timescales, are due not only to the
high specific star formation rates but also a limited molecular
gas reservoir.

\section{Conclusions}\label{s:conclusions}
We have observed a sample of 20 local LCBGs in the three lowest 
rotational transitions of CO. 14 galaxies were detected, including
all with blue luminosities $L_B>7\times 10^9~L_{\odot}$.
Using a Galactic conversion factor between CO intensity and H$_2$ column
density, we find H$_2$ masses
ranging from $6.6\times 10^6$ to $2.7\times 10^9~M_\odot$
with a median of $1.8\times 10^8~M_\odot$, including upper limits. 
The molecular to atomic hydrogen mass fraction is low,
typically $\lesssim 10\%$, which may be due to the 
shallow gravitational potential of these galaxies.

The line ratios of the three lowest CO transitions are similar to those
found in other low mass star bursting galaxies. Using an LVG model,
we find that the \emph{average} gas conditions in local LCBGs are similar to 
those found in star forming regions in our Galaxy, and nuclear regions of 
other galaxies.

Star formation rates, inferred from IRAS fluxes, range from 0.4 to 14
$M_\odot$ year$^{-1}$ with a median of 1.5 $M_\odot$ year$^{-1}$.
When normalized by dynamical mass (i.e., the specific star formation rate),
this is much larger than normal spiral galaxies.
If stars form at a constant rate, the molecular gas will be
used up in $10-200$~Myr, similar to depletion timescales in ULIRGs
and local strongly-interacting spirals.
We emphasise that nearly half the local sample
of local LCBGs have optical companions at similar velocities (Paper I).
Many have disturbed morphologies and some are actively interacting.
It is likely that interactions are even more common for the higher
redshift sample because of the higher space
density of galaxies \citep{2000A&G....41b..10E}.
Understanding the possible end-states resulting from such mergers requires
higher resolution (interferometric) observations and is the subject
of a future paper.

\bigskip
\acknowledgements
We thank Phil Solomon for encouraging us to compare our results with
his ULIRG studies, Dave Sanders and Bob Joseph for helpful discussions
about SEDs and SFRs, and Leslie Sage for providing copies of his CO
observations of Mrk~297.  
We also thank Harold Butner, Baltasar Vila-Vilaro, Herv\'e Aussel,
Axel Weiss, and the staff at the IRAM 30~m, JCMT and CSO
for their help with the observations and data calibration. 
D. J. P. acknowledges generous support from NSF MPS Distinguished
Research Fellowship grant AST-0104439.

Funding for the creation and distribution of the SDSS Archive has been
provided by the Alfred P. Sloan Foundation, the Participating
Institutions, the National Aeronautics and Space Administration, the
National Science Foundation, the U.S. Department of Energy, the Japanese
Monbukagakusho, and the Max Planck Society. The SDSS Web site is
http://www.sdss.org/.
The SDSS is managed by the Astrophysical Research Consortium (ARC) for
the Participating Institutions. The Participating Institutions are The
University of Chicago, Fermilab, the Institute for Advanced Study, the
Japan Participation Group, The Johns Hopkins University, Los Alamos
National Laboratory, the Max-Planck-Institute for Astronomy (MPIA), the
Max-Planck-Institute for Astrophysics (MPA), New Mexico State University,
University of Pittsburgh, Princeton University, the United States Naval
Observatory, and the University of Washington.
This research has also made use of the NASA/IPAC Infrared Science 
Archive, which is operated by the 
Jet Propulsion Laboratory, California Institute of Technology, under 
contract with the National Aeronautics and Space Administration.
We have made use of the NASA/IPAC Extragalactic Database (NED) 
which is operated by the Jet Propulsion Laboratory, 
California Institute of Technology, under contract with NASA
(http://nedwww.ipac.caltech.edu/).  

\bibliography{ms} 
%%

%%
% Table of basic properties
%%
\begin{deluxetable}{ccccccc}
\tabletypesize{\small}
\tablecaption{Properties of local LCBG Sample
\label{t:basic_prop}}
\tablewidth{0pt}
\tablehead{
\colhead{Source}	&
\colhead{D$_{OPT}$\tablenotemark{a}}	&
\colhead{D$_{25}$\tablenotemark{b}}	&
\colhead{L$_B$\tablenotemark{c}}		&
\colhead{L$_{IR}$}	&
\colhead{SFR}		&
\colhead{Companions}	\\
\colhead{}		&
\colhead{(Mpc)}		&
\colhead{(arc sec)}	&
\colhead{(10$^9$ L$_{B\odot}$)}	&
\colhead{(10$^9$ L$_\odot$)}	&
\colhead{(M$_\odot$ year$^{-1}$)}	&
\colhead{within 5$^\prime$}		
}
\startdata
Mrk 297	&	67	&	49	&	39	&	47	&	14	&	merger	\\
Mrk 314	&	30	&	52	&	3.9	&	1.6	&	0.50	&		\\
Mrk 325	&	49	&	90	&	16	&	18	&	5.5	&		\\
Mrk 538	&	40	&	114	&	17	&	24	&	7.3	&	$\checkmark$	\\
SDSS J011932.95+145219.0	&	59	&	40	&	5.6	&	3.7	&	1.1	&		\\
SDSS J021808.75$-$075718.0	&	69	&	29	&	5.2	&	2.4	&	0.74	&		\\
SDSS J022211.96$-$083036.2	&	67	&	26	&	4.3	&		&		&		\\
SDSS J072849.75+353255.2	&	56	&	40	&	5.6	&	3.4	&	1.0	&		\\
SDSS J083431.70+013957.9	&	59	&	44	&	6.8	&	6.5	&	2.0	&	$\checkmark$	\\
SDSS J090433.53+513651.1	&	68	&	44	&	12	&	8.5	&	2.6	&		\\
SDSS J091139.74+463823.0	&	61	&	38	&	8.2	&	5.1	&	1.5	&		\\
SDSS J093410.52+001430.2	&	70	&	41	&	13	&	22	&	6.8	&	$\checkmark$	\\
SDSS J093635.36+010659.8	&	71	&	42	&	6.8	&	4.4	&	1.3	&	$\checkmark$	\\
SDSS J094302.60$-$021508.9	&	68	&	29	&	8.2	&	6.1	&	1.8	&		\\
SDSS J111836.35+631650.4	&	46	&	28	&	4.3	&	1.3	&	0.38	&		\\
SDSS J123440.89+031925.1	&	67	&	42	&	7.4	&	4.9	&	1.5	&		\\
SDSS J131949.93+520341.1	&	67	&	33	&	4.7	&	3.7	&	1.10	&	$\checkmark$	\\
SDSS J140203.52+095545.6	&	61	&	59	&	12	&	16	&	4.7	&		\\
SDSS J150748.33+551108.6	&	48	&	67	&	5.6	&	1.4	&	0.42	&		\\
SDSS J231736.39+140004.3	&	63	&	48	&	8.2	&	11	&	3.4	&		\\
\enddata
\tablenotetext{a}{~Distance from optical redshifts, see Paper I.}
\tablenotetext{b}{~Isophotal diameter at the limiting surface brightness of 25 B-magnitudes arc sec$^{-2}$, from NED.}
\tablenotetext{c}{~From Paper I.}
\end{deluxetable}

%%
% Table of CO measurements
%%
\begin{deluxetable}{ccccccccc}
\rotate
\tabletypesize{\footnotesize}
\tablecaption{CO Line Measurements and Derived Properties
\label{t:co_line_all}}
\tablewidth{0pt}
\tablehead{
\colhead{Source}		&
\colhead{$V_\odot$}		&
\colhead{$\sigma$}		&
\colhead{I$_{CO}$~(J=1$-$0)}	&
\colhead{I$_{CO}$~(J=2$-$1)}    &
\colhead{I$_{CO}$~(J=2$-$1)}	&
\colhead{I$_{CO}$~(J=3$-$2)}	&
\colhead{M(H$_2$)}		&
\colhead{$\tau$(H$_2$)}		\\
\colhead{}		&
\colhead{}	&
\colhead{}	&
\colhead{$\theta$ = 21$^\prime$$^\prime$}	&
\colhead{$\theta$ = 11$^\prime$$^\prime$}	&
\colhead{$\theta$ = 21$^\prime$$^\prime$}	&
\colhead{$\theta$ = 21$^\prime$$^\prime$}	&
\colhead{}	&
\colhead{}	\\
\colhead{}		&
\colhead{(km s$^{-1}$)}	&
\colhead{(km s$^{-1}$)}	&
\colhead{(K km s$^{-1}$)}	&
\colhead{(K km s$^{-1}$)}	&
\colhead{(K km s$^{-1}$)}	&
\colhead{(K km s$^{-1}$)}	&
\colhead{(10$^8$ M$_\odot$)}	&
\colhead{(10$^6$ years)}
}
\startdata
Mrk 297	&	4728	$\pm$	2	&	49	$\pm$	1	&		24	$\pm$	0.63	&		29	$\pm$	0.84	&		20	$\pm$	0.38	&		5.7	$\pm$	0.79	&		27	$\pm$	0.7	&		193	\\
Mrk 314	&	2105	$\pm$	4\tablenotemark{a}	&	37	$\pm$	2\tablenotemark{a}	&	$<$	0.33			&		1.2	$\pm$	0.11	&	$<$	1.0			&					&		0.066	$\pm$	0.0061	&		13	\\
Mrk 325	&	3423	$\pm$	1	&	22	$\pm$	1	&		2.8	$\pm$	0.081	&		4.1	$\pm$	0.13	&		3.2	$\pm$	0.49	&		2.5	$\pm$	0.17	&		1.6	$\pm$	0.05	&		29	\\
Mrk 538	&	2799	$\pm$	2	&	51	$\pm$	1	&		6.8	$\pm$	0.21	&		18	$\pm$	0.33	&		11	$\pm$	0.42	&		8.2	$\pm$	0.58	&		2.6	$\pm$	0.08	&		36	\\
SDSSJ0119+1452	&				&				&					&					&	$<$	1.3			&					&	$<$	1.1			&	$<$	100	\\
SDSSJ0218$-$0757	&				&				&					&					&	$<$	1.3			&					&	$<$	1.6			&	$<$	216	\\
SDSSJ0222$-$0830	&				&				&					&					&	$<$	1.4			&					&	$<$	1.6			&			\\
SDSSJ0728+3532	&				&				&					&					&	$<$	1.2			&					&	$<$	0.91			&	$<$	91	\\
SDSSJ0834+0139	&				&				&	$<$	0.62			&	$<$	1.2			&	$<$	0.92			&	$<$	1.1			&	$<$	0.52			&	$<$	26	\\
SDSSJ0904+5136	&	4778	$\pm$	2	&	41	$\pm$	1	&		4.6	$\pm$	0.21	&		4.2	$\pm$	0.50	&		4.1	$\pm$	0.28	&		2.9	$\pm$	0.68	&		5.1	$\pm$	0.23	&		196	\\
SDSSJ0911+4636	&	4301	$\pm$	4	&	41	$\pm$	2	&		2.2	$\pm$	0.16	&		1.7	$\pm$	0.25	&		2.0	$\pm$	0.31	&		1.4	$\pm$	0.39	&		2.0	$\pm$	0.15	&		133	\\
SDSSJ0934+0014	&	4901	$\pm$	1	&	59	$\pm$	1	&		4.5	$\pm$	0.071	&		8.5	$\pm$	0.14	&		5.3	$\pm$	0.59	&		1.6	$\pm$	0.30	&		5.3	$\pm$	0.08	&		78	\\
SDSSJ0936+0106	&	4913	$\pm$	8	&	57	$\pm$	4	&		1.7	$\pm$	0.21	&	$<$	2.3			&	$<$	0.94			&					&		2.0	$\pm$	0.26	&		154	\\
SDSSJ0943$-$0215	&	4815	$\pm$	4	&	40	$\pm$	2	&		2.7	$\pm$	0.21	&		5.3	$\pm$	0.63	&		3.1	$\pm$	0.27	&		2.7	$\pm$	0.43	&		3.0	$\pm$	0.23	&		167	\\
SDSSJ1118+6316	&	3234	$\pm$	10	&	33	$\pm$	3	&		0.7	$\pm$	0.18	&		3.8	$\pm$	0.66	&		0.8	$\pm$	0.14	&	$<$	1.2			&		0.38	$\pm$	0.094	&		100	\\
SDSSJ1234+0139	&	4687	$\pm$	8	&	59	$\pm$	4	&		2.0	$\pm$	0.24	&	$<$	2.2			&		2.2	$\pm$	0.26	&	$<$	2.2			&		2.2	$\pm$	0.26	&		147	\\
SDSSJ1319+5203	&				&				&					&					&	$<$	1.0			&					&	$<$	1.1			&	$<$	100	\\
SDSSJ1402+0955	&	4259	$\pm$	4	&	43	$\pm$	2	&		2.9	$\pm$	0.22	&		4.8	$\pm$	0.45	&		3.1	$\pm$	0.25	&					&		2.6	$\pm$	0.20	&		55	\\
SDSSJ1507+5511	&	3366	$\pm$	12\tablenotemark{b}	&	13	$\pm$	3\tablenotemark{b}	&					&					&		0.5	$\pm$	0.19	&					&		0.29	$\pm$	0.11	&		69	\\
SDSSJ2317+1400	&	4434	$\pm$	2	&	32	$\pm$	1	&		6.7	$\pm$	0.27	&		13	$\pm$	0.75	&		7.1	$\pm$	0.44	&		4.9	$\pm$	0.47	&		6.5	$\pm$	0.26	&		191	\\
\enddata
\tablenotetext{a}{~from J=2$-$1 $\theta$ = 11$^\prime$$^\prime$}
\tablenotetext{b}{~from J=2$-$1 $\theta$ = 21$^\prime$$^\prime$}
\end{deluxetable}

%%
% Table of CO line ratios
%%
\begin{deluxetable}{cccc}
\tabletypesize{\small}
\tablecaption{CO Integrated Intensity Ratios (all $\theta$ = 21$^\prime$$^\prime$)
\label{t:line_ratios}}
\tablewidth{0pt}
\tablehead{
\colhead{Source}		&
\colhead{$\mathcal{R}_{21}$}	&
\colhead{$\mathcal{R}_{31}$}	&
\colhead{$\mathcal{R}_{32}$}	
}
\startdata
Mrk 297	&	0.83	$\pm$	0.026	&	0.24	$\pm$	0.033	&	0.29	$\pm$	0.041	\\
Mrk 325	&	1.1	$\pm$	0.14	&	0.89	$\pm$	0.068	&	0.78	$\pm$	0.060	\\
Mrk 538	&	1.6	$\pm$	0.079	&	1.2	$\pm$	0.093	&	0.75	$\pm$	0.060	\\
SDSSJ0904+5136	&	0.89	$\pm$	0.074	&	0.63	$\pm$	0.15	&	0.71	$\pm$	0.17	\\
SDSSJ0911+4636	&	0.91	$\pm$	0.16	&	0.64	$\pm$	0.18	&	0.70	$\pm$	0.22	\\
SDSSJ0934+0014	&	1.2	$\pm$	0.13	&	0.36	$\pm$	0.067	&	0.30	$\pm$	0.067	\\
SDSSJ0943$-$0215	&	1.1	$\pm$	0.13	&	1.0	$\pm$	0.18	&	0.87	$\pm$	0.16	\\
SDSSJ1118+6316	&	1.1	$\pm$	0.33	&		&		\\
SDSSJ1234+0139	&	1.1	$\pm$	0.19	&		&		\\
SDSSJ1402+0955	&	1.1	$\pm$	0.12	&		&		\\
SDSSJ2317+1400	&	1.1	$\pm$	0.080	&	0.73	$\pm$	0.077	&	0.69	$\pm$	0.080	\\
\enddata
\end{deluxetable}

%%
% Table of LVG NCO used
%%
\begin{deluxetable}{cc}
\tabletypesize{\small}
\tablecaption{Inferred N$_{CO}$ used in LVG Models
\label{t:lvg}}
\tablewidth{0pt}
\tablehead{
\colhead{Source}				&
\colhead{N$_{CO}$}			\\
\colhead{}				&
\colhead{(cm$^{-2}$)}			
}
\startdata
Mrk 297	&	4.5	$\times$	10$^{17}$	\\
Mrk 325	&	5.2	$\times$	10$^{16}$	\\
Mrk 538	&	1.2	$\times$	10$^{17}$	\\
SDSSJ0904+5136	&	8.3	$\times$	10$^{16}$	\\
SDSSJ0911+4636	&	3.4	$\times$	10$^{16}$	\\
SDSSJ0934+0014	&	8.1	$\times$	10$^{16}$	\\
SDSSJ0943$-$0215 &	4.9	$\times$	10$^{16}$	\\
SDSSJ2317+1400	&	1.2	$\times$	10$^{17}$	\\
\enddata
\end{deluxetable}

\newpage

\end{document}